\documentclass[
  aps,
  prd,
  reprint,
  onecolumn,
  superscriptaddress,
  amsmath,
  amssymb,
  longbibliography,
  floatfix,
  nofootinbib,
  eqsecnum
]{revtex4-2}

\usepackage{mathtools}
\usepackage{bm}
\usepackage{graphicx}
\usepackage{xcolor}

\usepackage[
  colorlinks=true,
  linkcolor=blue,
  citecolor=blue,
  urlcolor=blue,
  bookmarksnumbered=true
]{hyperref}

\usepackage{orcidlink}

\newcommand{\ii}{\mathrm{i}}
\newcommand{\dd}{\mathrm{d}}

\newcommand{\Gammacoh}{\Gamma_{\mathrm{coh}}}

\begin{document}

\title{Pulsar Timing Correlations Induced by Stochastic Spin-2 Ultralight Dark Matter}

\author{Hao-Xuan Shi}
\affiliation{Center for Gravitation and Cosmology, College of Physical Science and Technology, Yangzhou University, Yangzhou, 225009, China}

\author{Yu-Mei Wu\orcidlink{0000-0002-9247-5155}}
\email[Corresponding author:]{wuyumei@yzu.edu.cn} 
\affiliation{Center for Gravitation and Cosmology, College of Physical Science and Technology, Yangzhou University, Yangzhou, 225009, China}

\author{Chang Liu}
\email[Corresponding author:]{liuchang@yzu.edu.cn}
\affiliation{Center for Gravitation and Cosmology, College of Physical Science and Technology, Yangzhou University, Yangzhou, 225009, China}


\begin{abstract}
Ultralight dark matter (ULDM) can generate time-dependent gravitational perturbations, providing a potential signal for pulsar timing arrays (PTAs). In Galactic halos, the velocity dispersion of ULDM gives rise to finite spatial coherence, which influences the correlations between pulsar timing signals. In this work, we derive the timing-residual cross-correlations induced by rapid metric oscillations of spin-2 ULDM, allowing for general polarization statistics. Under equipartition among the five polarization modes, the intrinsic angular kernel contains monopole and quadrupole components and exhibits an angular profile closely resembling that of spin-1 ULDM, suggesting a potential degeneracy between the two ULDM candidates in PTA observations.
\end{abstract}

\maketitle

\section{Introduction}
\label{sec:intro}

Astrophysical and cosmological observations provide compelling evidence for dark matter, which accounts for most of the matter in the Universe~\cite{Planck2018}, although its microscopic nature remains unknown~\cite{Bertone2005}. The standard $\Lambda$CDM model, in which cold dark matter (CDM) plays a central role, successfully describes the cosmic microwave background and the large-scale structure of the Universe~\cite{Springel2006,Planck2018}. Despite this success, long-standing challenges associated with the CDM paradigm remain on small scales, including the core--cusp and missing-satellites problems~\cite{Bullock2017}. These challenges motivate alternatives to CDM, including ULDM, whose macroscopic de Broglie wavelength can make wave effects relevant to structure formation on astrophysical scales~\cite{Hu2000,Hui2017}, while its behavior can remain approximately CDM-like on sufficiently large scales~\cite{Ferreira2021}. The large occupation number of ULDM in Galactic halos justifies a classical-field description~\cite{Hui2017,Ferreira2021,Cheong:2024ose}. The wave-like nature of ULDM can produce observable effects over a wide range of frequencies, motivating searches using pulsar timing~\cite{KhmelnitskyRubakov2014}, laser interferometry~\cite{AokiSoda2017,Yu2024Laser,Zhang2025SpaceGW}, and astrometry~\cite{DrorVerner2025,Kim2024}. ULDM can also be probed through its modulation of astrophysical gravitational-wave (GW) signals~\cite{Brax2024,BlasGasparottoVicente2025}.

Among these observational approaches, pulsar timing arrays (PTAs), which monitor the pulse arrival times of millisecond pulsars~\cite{FosterBacker1990,BurkeSpolaor2019}, can probe ULDM with masses around $10^{-24}$--$10^{-22}\,\mathrm{eV}$~\cite{Porayko2018,PPTA:2022eul,EPTA2023}. The ULDM stress-energy tensor can source time-dependent metric perturbations, which induce fractional frequency shifts in propagating radio pulses and produce observable timing residuals~\cite{KhmelnitskyRubakov2014,Detweiler1979,Sun2022,Wu2023}. This gravitational effect has been investigated for ULDM with different spins~\cite{KhmelnitskyRubakov2014,Nomura2020,Wu2023,Unal2024}. In particular, spin-2 ULDM can arise in ghost-free bimetric gravity, which contains a massless spin-2 mode mediating ordinary gravitational interactions and a massive spin-2 mode that can serve as a dark matter candidate~\cite{Babichev:2016hir,Babichev:2016bxi,Hassan:2011zd,Marzola2018}. The massive mode has five physical polarization degrees of freedom, potentially giving rise to a rich geometric structure in its gravitational response. Indeed, previous studies have shown that the angular dependence of the timing signals can differ among spin-0 (scalar), spin-1 (vector), and spin-2 (tensor) ULDM, reflecting their different polarization structures~\cite{Nomura2020,Wu2023,Omiya:2023bio,Nomura:2025jur}. In addition to gravitational effects, ULDM can produce timing signals through direct couplings to Standard Model fields~\cite{Smarra2024,WuHuang2025,Xia2023}. Such couplings have also been considered in studies of angular correlations induced by spin-2 ULDM~\cite{Cai2024}.

Many earlier studies adopted simplified descriptions of the ULDM field, assuming monochromatic oscillations and considering two limiting cases of spatial coherence. In the fully correlated limit, the Earth and pulsars share the same ULDM field realization, whereas in the fully uncorrelated limit, their local fields are statistically independent. However, in a virialized Galactic halo, ULDM is more generally described as a superposition of nonrelativistic waves with different velocities and phases~\cite{Hui2017,KimMitridate2024,Boddy2025,DrorWei2025}. The interference between these waves leads to a finite spatial coherence length $\ell_{\rm coh}\sim(mv_0)^{-1}$, where $m$ is the ULDM mass and $v_0$ characterizes the velocity dispersion of dark matter in the Galactic halo~\cite{Hui2017,Boddy2025,DrorWei2025}. For part of the ULDM mass range accessible to PTAs, this coherence length can be comparable to Earth--pulsar distances, so the ULDM fields at the Earth and pulsar locations are generally partially correlated. A stochastic-field treatment that retains finite spatial correlations is therefore needed to characterize the timing signals more completely.

Recent studies have investigated the timing correlations induced by stochastic spin-0 and spin-1 ULDM, accounting for the velocity distribution and finite spatial coherence~\cite{KimMitridate2024,Boddy2025,DrorWei2025,Foster:2026mvs}. Because the stress-energy tensor is quadratic in the ULDM field, products of field modes generate sum- and difference-frequency contributions, corresponding to the conventional fast mode with angular frequency $\omega_{\rm F}\simeq 2m$ and the more recently identified slow mode with characteristic frequency $\omega_{\rm S}\sim mv_0^2$, respectively. The slow mode probes higher ULDM masses around $10^{-18}$--$10^{-16}\,\mathrm{eV}$~\cite{KimMitridate2024,DrorWei2025}. For spin-1 ULDM, existing stochastic calculations have adopted polarization equipartition among its three physical modes~\cite{DrorWei2025}.

For spin-2 ULDM, previous calculations of its gravitational timing response have largely relied on coherent-field descriptions~\cite{Wu2023}. In this work, we extend the analysis to a stochastic spin-2 field. We focus on the fast mode and derive the corresponding metric perturbations and timing-residual cross-correlations, accounting for finite spatial coherence and general polarization statistics among the five physical modes. Under five-polarization equipartition, we obtain an intrinsic angular kernel consisting of monopole and quadrupole components. Comparing this kernel with that of spin-1 ULDM under three-polarization equipartition, we find remarkably similar angular profiles, suggesting a potential degeneracy between the two spin sectors in PTA observations.

The remainder of this paper is organized as follows. Section~\ref{sec:field_metric} develops the field model and derives the corresponding metric perturbations, while Section~\ref{sec:pta_cov} derives the timing-residual cross-correlations. Section~\ref{sec:discussion} discusses the angular correlations and compares the spin-2 and spin-1 cases. Section~\ref{sec:conclusion} summarizes our findings.

We use natural units $c=\hbar=1$ and the metric signature $(-,+,+,+)$. Greek indices $\mu,\nu,\ldots$ denote spacetime components, while Latin indices $i,j,\ldots$ denote spatial components. Repeated indices are summed over, and Newton's constant $G$ is kept explicit.



\section{Stochastic Spin-2 ULDM Field and Fast-Mode Gravitational Response}
\label{sec:field_metric}

In this section, we begin by modeling spin-2 ULDM as a stochastic superposition of nonrelativistic field modes and then derive the gravitational response sourced by the sum-frequency components of its energy-momentum tensor.

\subsection{Stochastic Spin-2 Field}
\label{subsec:stochastic_spin2}

We describe the massive spin-2 field $M_{\mu\nu}$ propagating on an effective background metric $g_{\mu\nu}$ using the quadratic Fierz--Pauli action~\cite{FierzPauli1939,Babichev:2016hir,Babichev:2016bxi},
\begin{equation}
S_{\rm FP}^{(2)}=\int\dd^4x\,\sqrt{-g}\,\mathcal L_{\rm FP}^{(2)},
\label{eq:FP_action}
\end{equation}
with
\begin{equation}
\begin{aligned}
\mathcal L_{\rm FP}^{(2)}={}&
-\frac18\nabla_\lambda M_{\mu\nu}\nabla^\lambda M^{\mu\nu}
+\frac14\nabla_\lambda M_{\mu\nu}\nabla^\nu M^{\mu\lambda}
-\frac14\nabla_\mu M^{\mu\nu}\nabla_\nu M
+\frac18\nabla_\lambda M\nabla^\lambda M\\
&-\frac18m^2\left(M_{\mu\nu}M^{\mu\nu}-M^2\right),
\end{aligned}
\label{eq:FP_Lagrangian}
\end{equation}
where $M\equiv M^\mu{}_{\mu}$. On Galactic scales, we neglect cosmological expansion and
work on a flat background $g_{\mu\nu}\simeq\eta_{\mu\nu}$.
Varying the Fierz--Pauli action with respect to
$M_{\mu\nu}$ gives
\begin{equation}
\begin{aligned}
(\Box-m^2)M_{\mu\nu}
-\partial_\mu\partial^\rho M_{\nu\rho}
-\partial_\nu\partial^\rho M_{\mu\rho}
+\partial_\mu\partial_\nu M+\eta_{\mu\nu}
\left(
\partial^\rho\partial^\sigma M_{\rho\sigma}
-\Box M+m^2M
\right)=0,
\end{aligned}
\label{eq:FP_field_equation}
\end{equation}
Taking the divergence and trace of this equation
yields, for $m\ne0$,
\begin{equation}
M=0,\qquad
\partial^\mu M_{\mu\nu}=0,\qquad
(\Box-m^2)M_{\mu\nu}=0.
\label{eq:FP_constraints}
\end{equation}

For a Fourier mode with wave vector $\bm k$, the dispersion relation is $\omega_{\bm k}=\sqrt{m^2+k^2}$. In a virialized Galactic halo, $|\bm k|/m\sim v\sim v_0\sim 10^{-3}$, so $\omega_{\bm k}=m+k^2/(2m)+\mathcal O(mv^4)=m+\mathcal O(mv^2)$~\cite{Hui2017,Boddy2025,DrorWei2025}. Transversality gives $M_{0i}=\mathcal O(vM_{ij})$ and $M_{00}=\mathcal O(v^2M_{ij})$; together with $M=0$, this implies $M_{ii}=\mathcal O(v^2M_{ij})$. To leading nonrelativistic order, the five degrees of freedom are thus described by a spatial symmetric trace-free (STF) tensor $M_{ij}$~\cite{Marzola2018,Wu2023}. We expand it as
\begin{equation}
M_{ij}(t,\bm x)=
\sum_{A=1}^{5}\int\frac{\dd^3k}{(2\pi)^3}
M_A(\bm k)E^A_{ij}
 e^{-\ii\omega_{\bm k}t+\ii\bm k\cdot\bm x}
+\mathrm{c.c.},
\label{eq:random_field}
\end{equation}
where $E^A_{ij}$ form a fixed orthonormal STF basis and $M_A(\bm k)$ are complex random mode amplitudes.

We assume invariance of their joint distribution under a common phase rotation $M_A(\bm k)\to e^{\ii\phi}M_A(\bm k)$. This implies $\langle M_A(\bm k)\rangle=0$ and $\langle M_A(\bm k)M_B(\bm k')\rangle=0$, where $\langle\cdots\rangle$ denotes an ensemble average over stochastic field realizations~\cite{Boddy2025,DrorWei2025}. Statistical homogeneity gives the conjugated two-point function~\cite{AlonsoWolf2008,KuritaTakada2022}
\begin{equation}
\langle M_A(\bm k)M_B^*(\bm k')\rangle
=(2\pi)^3\delta^{(3)}(\bm k-\bm k')\mathcal P_{AB}(\bm k).
\label{eq:general_power_spectrum}
\end{equation}
To parameterize the momentum and polarization statistics independently, we adopt the separable model
\begin{equation}
\mathcal P_{AB}(\bm k)=P_M f_k(\bm k)W_{AB},
\label{eq:separable_power_spectrum}
\end{equation}
where $P_M$ fixes the overall power, $f_k(\bm k)$ describes the momentum distribution, and the Hermitian, positive-semidefinite matrix $W_{AB}$ describes the relative powers and cross-correlations of the five polarizations. We normalize them by
\begin{equation}
\int\frac{\dd^3k}{(2\pi)^3}f_k(\bm k)=1,
\qquad \operatorname{Tr}W=1.
\label{eq:statistical_normalization}
\end{equation}
The diagonal entries of $W$ specify the polarization powers, while its off-diagonal entries describe the  correlations between different polarizations. The polarization-equipartition case corresponds to $W_{AB}=\delta_{AB}/5$.

\subsection{Energy-Momentum Tensor and Frequency Components}
\label{subsec:energy_momentum}

The field sources gravity through its energy-momentum tensor, defined at quadratic order by the Hilbert prescription $T_{\mu\nu}=-(2/\sqrt{-g})\,\delta S_{\rm FP}^{(2)}/\delta g^{\mu\nu}$~\cite{Wu2023}. Using the free-field constraints and the nonrelativistic scaling $\partial_t\sim m$, $\partial_i\sim mv$, the leading energy density is
\begin{equation}
T_{00}=\frac18\left(\dot M_{ij}\dot M_{ij}
+\partial_kM_{ij}\partial_kM_{ij}+m^2M_{ij}M_{ij}\right).
\label{eq:T00_NR}
\end{equation}
Averaging over the rapid oscillations and the
stochastic ensemble gives the mean energy density,
\begin{equation}
\bar\rho\equiv\langle\overline{T_{00}}\rangle
=\frac12m^2P_M[1+\mathcal O(v^2)],
\label{eq:rho_PM}
\end{equation}
where the overbar denotes the oscillation average.
The leading spatial stress is
\begin{equation}
\begin{aligned}
T_{ij}={}&\frac18\delta_{ij}\left(\dot M_{k\ell}\dot M_{k\ell}
-m^2M_{k\ell}M_{k\ell}\right)
+\frac12\left(m^2M_{i\ell}M_{j\ell}
-\dot M_{i\ell}\dot M_{j\ell}\right)
+\mathcal O(\bar \rho v^2).
\end{aligned}
\label{eq:Tij_NR}
\end{equation}
The momentum density satisfies $T_{0i}=\mathcal O(\bar\rho v)$. 

Since $T_{\mu\nu}$ is quadratic in $M_{ij}$, pairs of modes produce sum- and difference-frequency components. The former have $\omega_{\bm k}+\omega_{\bm k'}=2m+\mathcal O(mv^2)$ (fast mode), while the latter have $\omega_{\bm k}-\omega_{\bm k'}=\mathcal O(mv^2)$ (slow mode)~\cite{KhmelnitskyRubakov2014,KimMitridate2024,DrorWei2025}. At leading nonrelativistic order, the sum-frequency terms survive in the oscillatory spatial stress, whereas the difference-frequency terms contribute to density fluctuations. 

Substituting Eq.~\eqref{eq:random_field} into Eq.~\eqref{eq:Tij_NR} and retaining the sum-frequency terms yields
\begin{equation}
\begin{aligned}
T_{ij}^{\rm F}(t,\bm x)={}&m^2\sum_{A,B}
\int\frac{\dd^3k\,\dd^3k'}{(2\pi)^6}
M_A(\bm k)M_B(\bm k')
\left[\frac12\left(E^A_{i\ell}E^B_{j\ell}
+E^A_{j\ell}E^B_{i\ell}\right)
-\frac14\delta_{ij}\delta^{AB}\right]\\
&\times e^{-\ii(\omega_{\bm k}+\omega_{\bm k'})t
+\ii(\bm k+\bm k')\cdot\bm x}
+\mathrm{c.c.}+\mathcal O(v^2\bar\rho).
\end{aligned}
\label{eq:Tij_fast}
\end{equation}
The fast-mode stress therefore retains the polarization-dependent tensor structure required for the gravitational response. 

The slow mode, corresponding to density fluctuations, produces a leading scalar gravitational response analogous to that of spin-0 and spin-1 ULDM~\cite{KimMitridate2024,DrorWei2025}. With a characteristic frequency $\omega_{\rm S}\sim mv^2$, it allows PTAs to probe higher ULDM masses, typically around $10^{-18}$--$10^{-16}\mathrm{eV}$. In the following, we focus on the fast-mode gravitational response.

\subsection{Fast-Mode Metric Perturbations}
\label{subsec:fast_metric}
\label{subsec:metric_response}

We perturb the flat metric as $g_{\mu\nu}=\eta_{\mu\nu}+h_{\mu\nu}$. The ULDM stress tensor sources the metric through the linearized Einstein equations,
\begin{equation}
\delta G_{\mu\nu}[h]=8\pi G T_{\mu\nu}.
\label{eq:linearized_einstein}
\end{equation}
For the fast branch, we use synchronous gauge, $h_{00}=h_{0i}=0$, and decompose the spatial perturbation as $h_{ij}=2s_{ij}-2\Psi\delta_{ij}$, where $s_{ii}=0$~\cite{Wu2023,DrorWei2025}. Using $\partial_i/\partial_t=\mathcal O(v)$, the spatial Einstein tensor at leading order reads
\begin{equation}
\delta G_{ij}\simeq \ddot s_{ij}+2\delta_{ij}\ddot\Psi.
\label{eq:einstein_spatial_fast}
\end{equation}
Decomposing the fast-mode stress into its trace and traceless parts, $T_{ij}^{\rm F}=\delta_{ij}T_{kk}^{\rm F}/3+\pi_{ij}^{\rm F}$ with $\pi_{ii}^{\rm F}=0$, gives
\begin{equation}
\ddot\Psi^{\rm F}=\frac{4\pi G}{3}T_{kk}^{\rm F},\qquad
\ddot s_{ij}^{\rm F}=8\pi G\pi_{ij}^{\rm F}.
\label{eq:einstein_fast_projections}
\end{equation}
For the sourced fast-mode response, $\partial_t^2\simeq-4m^2$ at leading order. Substituting Eq.~\eqref{eq:Tij_fast} yields
\begin{equation}
\begin{aligned}
h_{ij}^{\rm F}(t,\bm x)\simeq{}&-4\pi G\sum_{A,B}
\int\frac{\dd^3k\,\dd^3k'}{(2\pi)^6}
M_A(\bm k)M_B(\bm k')\mathcal R_{ij}^{AB}
 e^{-\ii(\omega_{\bm k}+\omega_{\bm k'})t
+\ii(\bm k+\bm k')\cdot\bm x}
+\mathrm{c.c.},
\end{aligned}
\label{eq:metric_fast}
\end{equation}
where
\begin{equation}
\mathcal R_{ij}^{AB}=\frac12\left(E^A_{i\ell}E^B_{j\ell}
+E^A_{j\ell}E^B_{i\ell}\right)
-\frac38\delta_{ij}\delta^{AB}.
\label{eq:Rij_response}
\end{equation}
At this order, $\mathcal R_{ij}^{AB}$ contains the polarization geometry and is independent of the mode momenta.

For the subsequent timing analysis, it is useful to separate the carrier frequency $2m$ from the slowly varying envelope,
\begin{equation}
h_{ij}^{\rm F}(t,\bm x)
=H_{ij}(t,\bm x)e^{-\ii 2mt}
+H_{ij}^*(t,\bm x)e^{\ii 2mt},
\label{eq:metric_fast_envelope}
\end{equation}
with
\begin{equation}
\begin{aligned}
H_{ij}(t,\bm x)={}&-4\pi G\sum_{A,B}
\int\frac{\dd^3k\,\dd^3k'}{(2\pi)^6}
M_A(\bm k)M_B(\bm k')\mathcal R_{ij}^{AB}
 e^{-\ii\delta\omega_{\bm k\bm k'}t
+\ii(\bm k+\bm k')\cdot\bm x},
\end{aligned}
\label{eq:metric_fast_envelope_definition}
\end{equation}
where $\delta\omega_{\bm k\bm k'}\equiv
\omega_{\bm k}+\omega_{\bm k'}-2m
=\mathcal O(mv^2)$.
The envelope varies on the kinetic-frequency scale $\mathcal O(mv^2)$ but remains part of the fast branch, distinct from the slow mode sourced by difference-frequency products.

\section{Pulsar Timing Response and Two-Point Statistics}
\label{sec:pta_cov}


For pulsar $a$, let $d_a$ be its distance from the Earth and $\hat{\bm n}_a$ the unit vector pointing toward it. With the Earth at $\bm x=0$, the pulsar is located at $\bm x_a=d_a\hat{\bm n}_a$. At leading order in the nonrelativistic expansion, the fractional frequency shift of the radio pulses is~\cite{Detweiler1979,Nomura2020,Wu2023,DrorWei2025}
\begin{equation}
 z_a^{\rm F}(t)\simeq\frac12\hat n_a^i\hat n_a^j
 \left[h_{ij}^{\rm F}(t,\bm 0)-h_{ij}^{\rm F}(t-d_a,\bm x_a)\right],
 \label{eq:redshift_fast}
\end{equation}
where the two terms are the Earth and pulsar contributions. The timing residual is $r_a^{\rm F}(t)=\int^t\dd t_1\,z_a^{\rm F}(t_1)$, with the integration constant absorbed into the timing model. The time integration of the fast oscillation introduces a factor $1/(2m)$, up to relative $\mathcal O(v^2)$ corrections. We now calculate the timing-residual cross-correlation between pulsars $a$ and $b$, $\langle r_a^{\rm F}(t)r_b^{\rm F}(t')\rangle$.


The fast-mode metric in Eq.~\eqref{eq:metric_fast_envelope_definition} is quadratic in the stochastic field amplitudes, so its two-point function involves their four-point statistics. We assume Gaussian mode amplitudes and apply Wick's theorem~\cite{Boddy2025,DrorWei2025}:
\begin{equation}
\begin{aligned}
&\langle M_A(\bm k)M_B(\bm k')M_C^*(\bm p)M_D^*(\bm p')\rangle=\langle M_A(\bm k)M_C^*(\bm p)\rangle
\langle M_B(\bm k')M_D^*(\bm p')\rangle+\langle M_A(\bm k)M_D^*(\bm p')\rangle
\langle M_B(\bm k')M_C^*(\bm p)\rangle.
\end{aligned}
\label{eq:wick_spin2}
\end{equation}
The pairing of two mode covariances in Eq.~\eqref{eq:general_power_spectrum}, together with the separable spectrum in Eq.~\eqref{eq:separable_power_spectrum}, then gives the factorized envelope correlation
\begin{equation}
 \left\langle H_{ij}(t,\bm x)H_{kl}^*(t',\bm x')\right\rangle
 =A_H\,\Gammacoh(\Delta t,\bm r)\,\mathcal K_{ij,kl}[W],
 \label{eq:metric_envelope_correlation}
\end{equation}
where $\Delta t\equiv t-t'$, $\bm r\equiv\bm x-\bm x'$, and
\begin{equation}
 A_H=(4\pi G)^2P_M^2=\frac{64\pi^2G^2\bar\rho^{\,2}}{m^4},
 \label{eq:AH_normalization}
\end{equation}
\begin{equation}
 \Gammacoh(\Delta t,\bm r)
 =\left[\int\frac{\dd^3k}{(2\pi)^3}f_k(\bm k)
 e^{-\ii(\omega_{\bm k}-m)\Delta t+\ii\bm k\cdot\bm r}\right]^2,
 \label{eq:Gamma_fast}
\end{equation}
\begin{equation}
 \mathcal K_{ij,kl}[W]=\sum_{A,B,C,D}
 \mathcal R_{ij}^{AB}\mathcal R_{kl}^{CD}
 \left(W_{AC}W_{BD}+W_{AD}W_{BC}\right).
 \label{eq:K_general}
\end{equation}
Thus, $\Gammacoh$ describes the spacetime coherence of the fast-mode envelope, whereas $\mathcal K_{ij,kl}[W]$ contains the polarization statistics and response geometry. Their factorization follows from the separable momentum--polarization distribution and the momentum-independent response tensor at leading nonrelativistic order.

Using $\bm k=m\bm v$ and $\omega_{\bm k}-m\simeq mv^2/2$, we express the coherence kernel in terms of the normalized velocity distribution $f_v(\bm v)=m^3f_k(m\bm v)/(2\pi)^3$:
\begin{equation}
 \Gammacoh(\Delta t,\bm r)
 =\left[\int\dd^3v\,f_v(\bm v)
 e^{-\ii mv^2\Delta t/2+\ii m\bm v\cdot\bm r}\right]^2.
 \label{eq:Gamma_velocity}
\end{equation}
For a shifted Maxwellian velocity distribution,
\begin{equation}
 f_v(\bm v)=\frac{1}{(2\pi v_0^2)^{3/2}}
 \exp\!\left[-\frac{|\bm v-\bar{\bm v}|^2}{2v_0^2}\right],
 \label{eq:Maxwell_velocity_distribution}
\end{equation}
where $\bar{\bm v}$ is the mean velocity and $v_0$ is the standard deviation of each Cartesian velocity component, following Refs.~\cite{Boddy2025,DrorWei2025}. The resulting coherence kernel is
\begin{equation}
\begin{aligned}
 \Gammacoh(\Delta t,\bm r)
 ={}&\frac{e^{-\ii m|\bar{\bm v}|^2\Delta t+2\ii m\bar{\bm v}\cdot\bm r}}
 {(1+\ii\Delta t/\tau_{\rm coh})^3}\times\exp\!\left[-\frac{|\bm r-\bar{\bm v}\Delta t|^2/\ell_{\rm coh}^2}
 {1+\ii\Delta t/\tau_{\rm coh}}\right],
\end{aligned}
 \label{eq:Gamma_MB}
\end{equation}
where $\ell_{\rm coh}=1/(mv_0)$ and $\tau_{\rm coh}=1/(mv_0^2)$. For the fast mode, the relevant time separations satisfy $|\Delta t|\ll\tau_{\rm coh}$, so the coherence kernel reduces to
\begin{equation}
 \Gammacoh(\Delta t,\bm r)\simeq
 \exp\!\left(-\frac{r^2}{\ell_{\rm coh}^2}
 +2\ii m\bar{\bm v}\cdot\bm r\right).
 \label{eq:Gamma_MB_short_time}
\end{equation}
Consequently, $\operatorname{Re}[e^{-\ii2m\Delta t}\Gammacoh]\simeq e^{-r^2/\ell_{\rm coh}^2}\cos[2m(\Delta t-\bar{\bm v}\cdot\bm r)]$, in agreement with the spacetime dependence of the fast-mode correlations in Ref.~\cite{DrorWei2025}. In the zero-mean-velocity limit used in Ref.~\cite{Boddy2025}, the kernel reduces to the Gaussian spatial suppression $e^{-r^2/\ell_{\rm coh}^2}$.

Since $\langle H_{ij}H_{kl}\rangle=0$ under the phase-invariant field ensemble, the real fast-mode metric satisfies
\begin{equation}
 \left\langle h_{ij}^{\rm F}(t,\bm x)h_{kl}^{\rm F}(t',\bm x')\right\rangle
 =2A_H\operatorname{Re}\!\left[
 e^{-\ii2m\Delta t}\Gammacoh(\Delta t,\bm r)\mathcal K_{ij,kl}[W]\right].
 \label{eq:metric_fast_correlation}
\end{equation}
Projecting this result onto the pulsar lines of sight, we define
\begin{equation}
 \mathcal G_{ab}[W]\equiv
 \hat n_a^i\hat n_a^j\hat n_b^k\hat n_b^l\mathcal K_{ij,kl}[W].
 \label{eq:Gab_general}
\end{equation}
Using Eq.~\eqref{eq:redshift_fast} and integrating the two redshifts over time gives the fast-mode timing-residual cross-correlation $C_{ab}(t,t')\equiv\langle r_a^{\rm F}(t)r_b^{\rm F}(t')\rangle$:
\begin{equation}
\begin{aligned}
 C_{ab}(t,t')\simeq{}
 \frac{8\pi^2G^2\bar\rho^{\,2}}{m^6}
 \operatorname{Re}\!\Bigg\{\mathcal G_{ab}[W]e^{-\ii2m\Delta t}
 \Big[&\Gammacoh(\Delta t,\bm 0)-e^{-\ii2md_b}\Gammacoh(\Delta t+d_b,-\bm x_b)
 -e^{\ii2md_a}\Gammacoh(\Delta t-d_a,\bm x_a)\\
 &+e^{\ii2m(d_a-d_b)}\Gammacoh(\Delta t-d_a+d_b,\bm x_a-\bm x_b)
 \Big]\Bigg\}.
\end{aligned}
 \label{eq:timing_residual_two_point}
\end{equation}
The four terms in brackets are the Earth--Earth, Earth--pulsar, pulsar--Earth, and pulsar--pulsar contributions. Within our leading-order separable model, they share the same polarization-geometric factor $\mathcal G_{ab}[W]$, while the endpoint phases and coherence kernel retain the dependence on pulsar distances and positions. The characteristic residual amplitude scales as $r_{\rm rms}\sim G\bar\rho/m^3$~\cite{Wu2023,DrorWei2025}.

Equation~\eqref{eq:timing_residual_two_point} applies to the general polarization matrix $W$ introduced in Sec.~\ref{sec:field_metric}. We examine the polarization-equipartition case and its intrinsic angular kernel, including the comparison with spin-1 ULDM, in Sec.~\ref{sec:discussion}.

\section{Discussion}
\label{sec:discussion}

As a natural benchmark, we consider rotationally invariant polarization statistics for spin-2 ULDM. Within our separable model, rotational invariance requires the polarization covariance matrix to be proportional to the identity, corresponding to polarization equipartition among the five STF components. Such an assumption is also physically motivated by the nonlinear gravitational evolution and subsequent virialization of Galactic dark matter halos, which may erase the memory of the initial polarization distribution, as discussed for spin-1 ULDM~\cite{DrorWei2025}. With $\operatorname{Tr}W=1$, we have
\begin{equation}
W_{AB}=\frac15\delta_{AB}. \label{eq:spin2_equipartition}
\end{equation}
Substituting this into Eq.~\eqref{eq:K_general} gives
\begin{equation}
 \mathcal K_{ij,kl}^{\rm eq}
 =\alpha\,\delta_{ij}\delta_{kl}
 +\beta(\delta_{ik}\delta_{jl}+\delta_{il}\delta_{jk}),
 \qquad
 \alpha=-\frac{107}{7200},\quad \beta=\frac{7}{300}.
 \label{eq:K_equal_isotropic}
\end{equation}
The coefficients follow from the STF completeness relation, as detailed in Appendix~\ref{app:polarization_algebra}. With $\mu_{ab}\equiv\hat{\bm n}_a\cdot\hat{\bm n}_b=\cos\xi_{ab}$, the intrinsic polarization-geometric kernel defined in Eq.~\eqref{eq:Gab_general} becomes
\begin{equation}
 \mathcal G_{ab}
 =-\frac{107}{7200}+\frac{7}{150}\mu_{ab}^2
 =\frac{1}{1440}P_0(\mu_{ab})+\frac{7}{225}P_2(\mu_{ab}).
 \label{eq:G_equal_legendre}
\end{equation}
where $P_0(\mu)=1$ and $P_2(\mu)=(3\mu^2-1)/2$ are the monopole and quadrupole Legendre polynomials, respectively. Thus, under five-polarization equipartition, the intrinsic spin-2 angular kernel contains only monopole and quadrupole components, with the latter dominant.

To compare the intrinsic angular kernels of different spin sectors at fixed ULDM mass and mean density, we factor out the common dimensional amplitude $\pi^2G^2\bar\rho^{2}/m^6$ from the fast-mode timing-residual covariance. The resulting dimensionless kernels for spin-0, spin-1, and spin-2 ULDM under polarization equipartition are
\begin{align}
\widehat{\mathcal G}
_{ab}^{(0)}=\frac14,\qquad
\widehat{\mathcal G}_{ab}^{(1)}=\frac{16\mu_{ab}^2-5}{36},\qquad 
\widehat{\mathcal G}_{ab}^{(2)}
=\frac{336\mu_{ab}^2-107}{900}.
\end{align}
The spin-0 and spin-1 results are taken from Ref.~\cite{Boddy2025} and \cite{DrorWei2025}, respectively, while the spin-2 result follows from Eq.~\eqref{eq:G_equal_legendre}.

\begin{figure}[htbp]
    \centering    \includegraphics[width=3.4in]{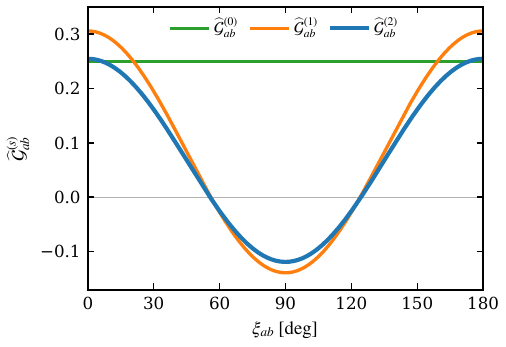}
    \caption{Dimensionless intrinsic angular kernels for spin-0 (green), spin-1 (orange), and spin-2 (blue) ULDM under polarization equipartition. The kernels are normalized by the common dimensional factor $\pi^2G^2\bar\rho^{,2}/m^6$.}
    \label{fig:spin012_angular}
\end{figure}

As illustrated in Fig.~\ref{fig:spin012_angular}, the spin-0 kernel is independent of angular separation, whereas the spin-1 and spin-2 kernels exhibit similar quadrupole-dominated angular dependence. In particular, they satisfy
\begin{equation}
\widehat{\mathcal G}_{ab}^{(2)}
=\frac{107}{125}\widehat{\mathcal G}_{ab}^{(1)}
-\frac{8}{1125}\mu_{ab}^2.
\end{equation}
The small correction to the rescaled spin-1 kernel highlights the close similarity of their angular profiles, despite their different overall amplitudes.

These results suggest that, if polarization equipartition is established during the evolution of Galactic dark matter halos, the similarity of these angular profiles may make spin-1 and spin-2 ULDM difficult to distinguish using PTA timing correlations, particularly since differences in their overall amplitudes can be absorbed into the inferred ULDM density.

For completeness, the slow-mode density-fluctuation correlations depend on the polarization matrix through $\operatorname{Tr}(W^2)$ under the Gaussian and separable statistical model adopted here. Under polarization equipartition, the resulting slow-mode timing-residual correlations have relative amplitudes $1$, $1/3$, and $1/5$ for spin-0, spin-1, and spin-2 ULDM, respectively, when compared at fixed mean density, mass, and velocity distribution. A short derivation is given in Appendix~\ref{app:slow_mode_covariance}.

\section{Conclusion}
\label{sec:conclusion}

We have investigated the stochastic gravitational timing signal induced by the fast mode of spin-2 ULDM. Modeling the nonrelativistic field as a stochastic superposition of five physical polarizations, we derived the metric perturbations and pulsar timing-residual correlations under Gaussian field statistics and a separable separable momentum-polarization power spectrum. The resulting covariance incorporates both Earth and pulsar terms, with the polarization geometry factorized from the spacetime coherence of the field. This framework extends previous coherent-field treatments by accounting for stochastic field statistics and finite spatial coherence.

Under five-polarization equipartition, the intrinsic angular kernel contains monopole and quadrupole components, with the quadrupole dominant. Comparing the spin-0, spin-1, and spin-2 angular kernels at fixed ULDM mass and mean density, we find that spin-1 and spin-2 ULDM exhibit remarkably similar angular profiles, despite their different overall amplitudes. This similarity is not a generic consequence of rotational invariance, which does not fix the relative weights of different multipoles. If polarization equipartition is established during the evolution of Galactic dark matter halos, these results suggest a potential degeneracy between the two ULDM candidates in PTA observations, as their amplitude differences can be absorbed into the inferred ULDM density.

For completeness, we have also examined the slow mode, whose leading gravitational response arises from density fluctuations. Under the Gaussian and separable statistical assumptions adopted here, its density correlations depend on the polarization covariance through $\operatorname{Tr}(W^2)$. Under polarization equipartition, this yields relative slow-mode timing-correlation amplitudes of $1$, $1/3$, and $1/5$ for spin-0, spin-1, and spin-2 ULDM, respectively.


\begin{acknowledgments}
 
YMW is supported by the National Natural Science Foundation of China (Grant No.~12505086).
CL is supported by the National Natural Science Foundation of China (Grant No.~12405074).

\end{acknowledgments}

\bibliography{references}

\appendix

\section{Spin-2 Polarization Algebra}
\label{app:polarization_algebra}

For the five-polarization equipartition model in Sec.~\ref{sec:discussion}, the STF basis obeys the completeness relation
\begin{equation}
\sum_{A=1}^{5}E^A_{ij}E^A_{kl}
=\frac12(\delta_{ik}\delta_{jl}+\delta_{il}\delta_{jk})
-\frac13\delta_{ij}\delta_{kl}.
\label{eq:app_STF_projector}
\end{equation}
Using the response tensor $\mathcal R_{ij}^{AB}$ in Eq.~\eqref{eq:Rij_response}, Eq.~\eqref{eq:K_general} reduces to
$\mathcal K_{ij,kl}^{\rm eq}=(2/25)\sum_{A,B}
\mathcal R_{ij}^{AB}\mathcal R_{kl}^{AB}$. The two independent contractions are
\begin{equation}
\mathcal K_{ii,kk}^{\rm eq}=\frac{1}{160},
\qquad
\mathcal K_{ij,ij}^{\rm eq}=\frac{113}{480}.
\label{eq:app_equal_contractions}
\end{equation}
For the isotropic tensor decomposition $\mathcal K_{ij,kl}^{\rm eq}
=\alpha\delta_{ij}\delta_{kl}
+\beta(\delta_{ik}\delta_{jl}+\delta_{il}\delta_{jk})$, these contractions give $9\alpha+6\beta=1/160$ and $3\alpha+12\beta=113/480$, hence
\begin{equation}
\alpha=-\frac{107}{7200},
\qquad \beta=\frac{7}{300}.
\label{eq:app_alpha_beta_result}
\end{equation}
These coefficients determine the intrinsic spin-2 angular kernel in Eq.~\eqref{eq:G_equal_legendre}.


\section{Slow-Mode Density Correlations}
\label{app:slow_mode_covariance}

The slow mode arises from difference-frequency products of the nonrelativistic spin-2 field, which generate density fluctuations and source a leading scalar gravitational response, as in spin-0 and spin-1 ULDM~\cite{KimMitridate2024,DrorWei2025}.
Defining $\delta\rho^{\rm S}=T_{00}^{\rm S}-\bar\rho$,
we obtain
\begin{equation}
\begin{aligned}
\delta\rho^{\rm S}(t,\bm x)
={}&\frac{m^2}{2}\sum_A
\int\frac{\dd^3k\,\dd^3k'}{(2\pi)^6}
M_A(\bm k)M_A^*(\bm k')\,
e^{-\ii(\omega_{\bm k}-\omega_{\bm k'})t
+\ii(\bm k-\bm k')\cdot\bm x}
-\bar\rho .
\end{aligned}
\label{eq:app_slow_density_fluctuation}
\end{equation}

For the Gaussian field statistics and separable
momentum--polarization spectrum introduced in
Sec.~\ref{sec:field_metric}, Wick's theorem gives
the connected density correlation
\begin{equation}
\begin{aligned}
&\left\langle
\delta\rho^{\rm S}(t,\bm x)
\delta\rho^{\rm S}(t',\bm x')
\right\rangle_{\rm c}=
\bar\rho^{\,2}\operatorname{Tr}(W^2)
\mathcal F_{\rm S}(\Delta t,\bm r),
\end{aligned}
\label{eq:app_slow_density_covariance}
\end{equation}
where $\Delta t=t-t'$, $\bm r=\bm x-\bm x'$, and
\begin{equation}
\mathcal F_{\rm S}(\Delta t,\bm r)
=
\left|
\int\frac{\dd^3k}{(2\pi)^3}
f_k(\bm k)
e^{-\ii(\omega_{\bm k}-m)\Delta t
+\ii\bm k\cdot\bm r}
\right|^2.
\label{eq:app_slow_coherence_general}
\end{equation}
Thus, the polarization dependence enters through
$\operatorname{Tr}(W^2)$, while the spacetime
dependence is determined by the momentum distribution.

To obtain the gravitational response, we use the
Newtonian-gauge metric
\begin{equation}
\dd s^2=
-(1+2\Phi^{\rm S})\dd t^2
+(1-2\Psi^{\rm S})\delta_{ij}
\dd x^i\dd x^j,
\label{eq:app_slow_metric}
\end{equation}
where $\Phi^{\rm S}$ and $\Psi^{\rm S}$ are
the scalar metric potentials induced by the
slow-mode density fluctuations.
At leading nonrelativistic order, the linearized
Einstein equations give
\begin{equation}
\nabla^2\Psi^{\rm S}
=4\pi G\,\delta\rho^{\rm S},
\qquad
\Phi^{\rm S}\simeq\Psi^{\rm S}.
\label{eq:app_slow_Poisson}
\end{equation}

The gravitational acceleration of a nonrelativistic
body is $\dot v^i\simeq-\partial_i\Phi^{\rm S}$.
For the slow mode, the leading pulsar redshift is
dominated by the relative motion of the Earth
and pulsar induced by this acceleration
\cite{KimMitridate2024,DrorWei2025}.
Using the pulsar geometry defined in
Sec.~\ref{sec:pta_cov}, we obtain
\begin{equation}
\begin{aligned}
z_a^{\rm S}(t)\simeq
\hat n_a^i\int_{t_0}^{t}\dd t_1\,
\Big[
\partial_i\Phi^{\rm S}(t_1,\bm0)
-\partial_i\Phi^{\rm S}
(t_1-d_a,\bm x_a)
\Big],
\end{aligned}
\label{eq:app_slow_redshift}
\end{equation}
up to constant velocity contributions absorbed
by the timing model.
The corresponding timing residual is
$r_a^{\rm S}(t)=\int^t\dd t_1\,z_a^{\rm S}(t_1)$.

Since the potential, redshift, and timing residual
are linear functionals of $\delta\rho^{\rm S}$
at this order, their connected two-point functions
inherit the polarization factor
$\operatorname{Tr}(W^2)$.
For $N$ polarization modes under equipartition,
$W_{AB}=\delta_{AB}/N$, giving
$\operatorname{Tr}(W^2)=1/N$.
Consequently, at fixed ULDM mass, mean density,
and velocity distribution, and with the same
leading-order timing-response prescription,
\begin{equation}
C_{ab,\mathrm{spin}\text{-}0}^{\rm S}:
C_{ab,\mathrm{spin}\text{-}1}^{\rm S}:
C_{ab,\mathrm{spin}\text{-}2}^{\rm S}
=
1:\frac13:\frac15.
\label{eq:app_slow_spin_comparison}
\end{equation}
For general polarization statistics, the
corresponding factor is $\operatorname{Tr}(W^2)$
rather than $1/5$.

\end{document}